\documentclass[prb,twocolumn]{revtex4}
\usepackage{graphicx,amsmath,amssymb,float,bm,times}

\newcommand{\Lambdak}{\bm{\Lambda}}
\newcommand{\lambdak}{\bm{\lambda}}
\newcommand{\ek}{\epsilon_{\mathbf{k}}}
\newcommand{\ekq}{\epsilon_{\mathbf{k-q}}}
\newcommand{\Ek}{E_{\mathbf{k}}}

\newcommand{\phik}{\varphi_{\mathbf{k}}}
\newcommand{\phikq}{\varphi_{{\mathbf{k}}-{\mathbf{q}}/2}}
\newcommand{\mb}[1]{{\mathbf{#1}}}
\newcommand{\sumk}{\sum_{\mathbf{k}}}

\newcommand{\uk}{u_{\mathbf{k}}}
\newcommand{\vk}{v_{\mathbf{k}}}

\begin{document}

\title{ The pseudogap state in superconductors: Extended Hartree
  approach to time-dependent Ginzburg-Landau Theory} 

\author{Jelena Stajic$^1$, Andrew Iyengar$^1$, Qijin Chen$^2$ and K.
  Levin$^1$}

\affiliation{$^1$ James Franck Institute and Department of Physics,
  University of Chicago, Chicago, Illinois 60637} 

\affiliation{$^2$ Department of Physics and Astronomy, Johns Hopkins
  University, Baltimore, Maryland 21218} 

\date{\today}

\begin{abstract}

  It is well known that conventional pairing fluctuation theory at the
  Hartree level leads to a normal state pseudogap in the fermionic
  spectrum.  This pseudogap arises from bosonic degrees of freedom which
  appear at $ T ^*$, slightly above the superconducting transition
  temperature $ T_c$.  Our goal is to extend this Hartree approximated
  scheme to arrive at a generalized mean field theory of pseudogapped
  superconductors for all temperatures $T$.  While an equivalent
  approach to the pseudogap has been derived elsewhere using a more
  formal Green's function decoupling scheme, in this paper we
  re-interpret this mean field theory and BCS theory as well, and
  demonstrate how they naturally relate to ideal Bose gas condensation.
  Here we recast the Hartree approximated Ginzburg-Landau self
  consistent equations in a T-matrix form, from which we infer that the
  condition that the pair propagator have zero chemical potential at all
  $T \le T_c$ is equivalent to a statement that the fermionic excitation
  gap satisfies the usual BCS gap equation.  This recasting makes it
  possible to consider arbitrarily strong attractive coupling, where
  bosonic degrees of freedom appear at $ T^*$ considerably above $T_c$.
  The implications for transport both above and below $T_c$ are
  discussed.  Below $T_c$ we find two types of contributions. Those
  associated with fermionic excitations have the usual BCS functional
  form.  That they depend on the magnitude of the excitation gap,
  nevertheless, leads to rather atypical transport properties in the
  strong coupling limit, where this gap (as distinct from the order
  parameter) is virtually $T$-independent. In addition, there are
  bosonic terms arising from non-condensed pairs whose transport
  properties are shown here to be reasonably well described by an
  effective time-dependent Ginzburg-Landau theory.
\end{abstract}
\maketitle

\section{Introduction}

In this paper we address the physics of the pseudogap state on the basis
of a beyond-BCS mean field theory.  Our approach represents a natural
extension to arbitrarily strong coupling constants, $g$, of 
time-dependent Ginzburg-Landau (TDGL) theory when the quartic terms are
treated at the Hartree approximation level.  We also explore the broken
symmetry phase, incorporating the effects of a Hartree approximation to
pairing (or equivalently, excitation-gap) fluctuations for $T < T_c$ in
a fashion which is necessarily compatible with their description above
$T_c$.  A central assumption in our work is that the ground state
belongs to the generalized BCS-like theory introduced by
Leggett\cite{Leggett}, and further analyzed by Nozieres and
Schmitt-Rink\cite{NSR}, and others.  In this way, one arrives at a
simple re-interpretation of BCS theory as a limiting case of a more
general theory of ideal Bose gas condensation.  In contrast to true Bose
systems, here, however, fermionic degrees of freedom are never absent:
the condition $\mu_{pair} = 0$, for $T \le T_c$ is equivalent to the
constraint imposed by the BCS gap equation on the excitation gap
$\Delta(T)$.

Several key points need to be made. The observation of a pseudogap ($T >
T_c$) deriving from Hartree approximated treatment of ``pairing"
fluctuations dates back to the early 70's, with the work of
Abeles\cite{Abeles}, Patton\cite{Patton1971,Patton} and Hassing and
Wilkins\cite{Wilkins}.  Pairing fluctuations can be alternatively viewed
as fluctuations in the magnitude of the fermionic excitation gap, that
is ``gap fluctuations". Most importantly, at the Hartree level, these
are to be distinguished from order parameter fluctuations, as will be
discussed in more detail below.  Indeed, bosonic degrees of freedom,
associated with fermion pairs, and the existence of a fermionic
excitation gap are two sides of the same coin. \textit{The gap or
  pseudogap in the fermionic excitation spectrum reflects the fact that
  extra energy is needed to break bosons, or fermion pairs, apart}.
When approached from above, $T_c$ is suppressed (relative to its strict
BCS counterpart) by the existence of this normal state pseudogap, as a
consequence of the lowering of the fermionic density of states near
$E_F$. When approached from below, one readily sees that the superfluid
density will not vanish at $T^*$, where the fermionic gap disappears,
but instead at a lower temperature $T_c$, as a consequence of additional
(beyond BCS) bosonic excitations of the condensate.

In conventional (\textit{i.e.,} long coherence length, $\xi$) three
dimensional superconductors, these non-condensed pairs are virtually
undetectable. However, there are two mechanisms for enhancing their
effects: reduced dimensionality, often with disorder [see the recent
review\cite{Varlamov_review} by Varlamov and Larkin which addresses
Gaussian pair fluctuations] and increased attractive coupling
constant\cite{NSR,Randeriareview,Chen2}, associated with short $\xi$.
If by ``the fluctuation regime" we mean the temperature range where
bosonic degrees of freedom are present, then in the second mechanism one
is effectively extending this regime up to $T^*$, which could be as much
as an order of magnitude higher than $T_c$.  In the vicinity of Bose
condensation there are divergences in various transport properties, but
even away from $T_c$, closer to $T^*$, bosonic contributions may
dominate over their fermionic counterparts, leading to highly
unconventional transport characteristics, as will be outlined here.

The Hartree approximation\cite{Marcelja,Grossmann,Wilkins} and its
T-matrix extension\cite{Patton1971,Patton} is a previously well
established route for obtaining a pseudogap state in homogeneous
superconductors\cite{Abeles}.  Throughout this paper we call the latter
``extended Hartree theory".  Nevertheless, there is considerable
emphasis in the current literature on approaches to the pseudogap based
on an alternative phase fluctuation scenario\cite{Emery}.  The contrast
between these two schemes should be emphasized.  In the former case one
is dealing with non-condensed bosons, which are associated with
fluctuations in the fermionic \textit{excitation gap} $\Delta(r,t)$, (say,
measured relative to the spatially uniform or mean field order parameter).  
In the latter case, one is dealing with fluctuations in the
phase of the complex \textit{order parameter} $\Delta_{sc}(r,t)$.  At a
beyond-BCS level these two channels are distinguishable ($\Delta \ne
\Delta_{sc}$), but, as in any mean field theory, the order parameter is
taken to be spatially uniform.  By contrast, the phase fluctuation
scheme is based on order parameter fluctuations around the strict BCS
state ($\Delta = \Delta_{sc}$).  Generally, order parameter fluctuation
approaches are appropriate when the separation between the mean field
onset temperature $T^*$ and the actual instability temperature $T_c$ is
small.  If this temperature difference is not small, it is more
appropriate to derive an improved mean field theory first\cite{Larkin}
and then append fluctuation effects, if they are called for.  This is
the philosophy of the present approach in which at a mean field level,
the separation between $T^*$ and $T_c$ can be arbitrarily tuned via the
size of the attractive coupling constant.

Finally, we emphasize a contrast between the present approach and
previous work by Nozieres and Schmitt-Rink\cite{NSR} and by Randeria and
co-workers\cite{Randeriareview}. Both groups studied the effects of
pre-formed pairs, or non-condensed bosons, presuming a BCS-like ground
state with arbitrary attractive coupling $g$. We will assume this ground
state as well. These other studies were conducted at the level of
Gaussian pair fluctuations, so that pseudogap effects were absent in the
pair propagator.  Moreover, an approximate form for the number equation
was assumed as well\cite{Serene}.  \textit{Essential to the physics we
  will be discussing, is a pseudogap in the fermionic spectrum, which
  helps to decouple the bosonic and fermionic degrees of freedom}.  This
pseudogap leads to relatively long-lived bosons, without the need to
invoke unphysically strong coupling. A gap in the fermionic spectrum
inhibits the decay of bosons into fermions, simply because there are
very few low energy fermionic states to decay into.  As a consequence,
bosonic degrees of freedom are evident over a much wider temperature
range, above and below $T_c$, once $g$ is beyond the weak coupling
regime.

In this paper, we make no contact with experiments in high temperature
superconductors. The goal here is to demonstrate the simple physics, as
well as the genesis of pair fluctuation approaches, as distinguished
from phase fluctuation schemes.  It is useful to emphasize that we are
focusing in the present paper on the nature of the superconductivity
(below $T_c$) and its implications above $T_c$. While we presume a
generalized BCS ground state (albeit, with arbitrary $g$ and self
consistent fermionic chemical potential), other ground states have been
contemplated in the cuprate literature which accommodate the physics of
the Mott insulator to varying degrees.  The strongest support for the
relevance of our viewpoint (which effectively sidesteps Mott effects) is
the anomalously short coherence length $\xi$.  This suggests a breakdown
of strict BCS physics, which, at the very least, needs to be understood
and characterized, on its own, independent of Mott effects.  To support
this breakdown is the highly non-BCS temperature dependence of the
measured\cite{arpesanl,arpesstanford} excitation gap $\Delta$ -- not so
different from that shown in Fig.~\ref{fig:schematic} below. Although
transport calculations and interpretations of transport data are often
based on a BCS ground state, rather little attention has been paid in
the past to the anomalous behavior of $\Delta(T)$ and its implications
for transport.  In this paper we address this omission.

\section{Above $T_c$ at weak coupling}
\label{sec:above_Tc_weak}
\subsection{Overview of Hartree approximated Ginzburg-Landau theory}

The Ginzburg-Landau (GL) free energy functional in momentum space is
given by \cite{Wilkins}
\begin{eqnarray}
F[\Psi]&=&\frac{N(0) V}{\beta ^2} \sum_{Q} |\Psi_{Q}|^2 (\epsilon +a
|\Omega _n| +\xi _1^2q^2)\nonumber \\ 
&&{}+\frac{1}{2 \beta^2} \sum_{Q_i} b_{Q_1Q_2Q_3}
\Psi^*_{Q_1}\Psi^*_{Q_2}\Psi_{Q_3}\Psi_{Q_1+Q_2-Q_3} \:,\hspace{4ex}
\label{free_energy}
\end{eqnarray} 
where $\Psi_{Q}$ are the Fourier components of the order parameter
$\Psi({\mathbf r},t)$, $Q=(i \Omega _n,{\bf q})$,
$\epsilon=\frac{T-T^*}{T^*}$, $a=\frac{\pi}{8 T}$, $\xi _1$ is the GL
coherence length, $T^*$ is the critical temperature when feedback effects from the quartic term are neglected, $\beta =1/T$
($k_B$ is set to 1) and $N(0)$ is the density of states at the Fermi level
in the normal state.  We approximate the quartic term so that only
paired terms are included in the last addend of Eq.~(\ref{free_energy})
leading to
\begin{eqnarray} 
\lefteqn{\frac{1}{2 \beta ^2}\sum_{123}
b_{123}\Psi^*_1\Psi^*_2\Psi_3\Psi_{1+2-3}}\nonumber\\
&\approx& \frac{1}{2 \beta ^2}\Big(\sum_{1\neq 2} b_{12}
|\Psi_1|^2|\Psi_2|^2+\sum_1 b_{11} |\Psi_1|^4\Big) \:.
\label{quartic_1}
\end{eqnarray}
That there is no factor of 2 in the first term on the right-hand side of the above expression reflects the fact that we use
Hartree rather than the Hartree-Fock approximation.  As found
elsewhere\cite{Wilkins}, $b_{ij} = b_{Q_iQ_j}$ can be approximated by
$b_0\delta_{\Omega_i0}\delta_{\Omega_j0}$ where $b_0=[N(0)V/\pi^2]
\frac{7}{8}\zeta(3)$.  To further simplify the quartic term, we apply
the mean field approximation, writing $|\Psi_{{\bf
    q}0}|^2=\langle|\Psi_{{\bf q}0}|^2\rangle +\delta|\Psi_{{\bf
    q}0}|^2$ and neglecting in Eq.~(\ref{quartic_1}) terms of order
$(\delta |\Psi_{{\bf q}0}|^2)^2$. This leads to
\begin{equation}
  \frac{b_0}{\beta^2}\sum_{\bf q} \Big(|\Psi_{{\bf
      q}0}|^2-\frac{1}{2}\langle|\Psi_{{\bf q}0}|^2\rangle\Big)\sum_{{\bf
      q}'}\langle|\Psi_{{\bf q}'0}|^2\rangle
\label{quartic_2}
\end{equation}
The contribution $\langle |\Psi_{{\bf q}0}|^2\rangle$ is determined
self-consistently via
\begin{equation}
\langle |\Psi_{{\bf q}0}|^2\rangle=\frac{\int D\Psi e^{-\beta F[\Psi]}
  |\Psi_{{\bf q}0}|^2}{\int D\Psi e^{-\beta F[\Psi]}} 
\label{psi_q}
\end{equation}
when we replace the quartic term in Eq.~(\ref{free_energy}) by
Eq.~(\ref{quartic_2}). It follows that
\begin{equation}
\langle |\Psi_{{\bf q}0}|^2\rangle=\frac{1}{N(0) V  T}\bigg[\epsilon +
\frac{b_0}{N(0)V}\sum_{{\bf q}'} \langle |\Psi_{{\bf q}'0}|^2\rangle 
+\xi _1^2 q^2\bigg]^{-1}. 
\label{self_cons_psi_q}
\end{equation}
If we sum Eq.~(\ref{self_cons_psi_q}) over ${\bf q}$ and identify
$\sum_{\bf q} \langle |\Psi_{{\bf q}0}|^2\rangle$ with $\beta^2 \Delta
^2$, we obtain a self-consistency equation for the energy "gap" (or
pseudogap) $\Delta$ above $T_c$
\begin{equation}
\beta ^2 \Delta ^2=\sum_{\bf q} \frac{1}{N(0) V T}\bigg[\epsilon + 
\frac{b_0}{N(0) V}\beta ^2 \Delta ^2
+\xi _1^2 q^2\bigg]^{-1},
\label{gap_GL}
\end{equation}
\begin{equation}
\beta ^2 \Delta ^2=\sum_{\bf q} \frac{1}{N(0) V T}\; \frac{1}{-
  \bar{\mu}_{pair}(T) +\xi _1^2 q^2} \;,
\label{gap_GL2}
\end{equation}
where
\begin{equation}
\bar{\mu}_{pair}(T)= -\epsilon- \frac{b_0}{N(0) V}\beta^2\Delta^2
\label{mu_def_GL}
\end{equation}
Note that the critical temperature is renormalized downward with respect
to $T^*$ and satisfies
\begin{equation}
\bar{\mu}_{pair}(T_c)=0
\label{tc_cond_GL}
\end{equation}

\subsection{Introduction to T-matrix : $T \approx T_c$,
  small $\Delta(T_c)$ } 

Equations (\ref{gap_GL2}), (\ref{mu_def_GL}), and (\ref{tc_cond_GL}) are
the central equations derived from the Hartree approximated
Ginzburg-Landau (GL) scheme. They describe how the excitation gap $
\Delta(T)$ and the quantity $\bar{\mu}_{pair}$ behaves above, but near $
T_c$. We now rewrite these equations using a T-matrix approach.

A central quantity in T-matrix schemes is
the pair susceptibility. Here we take this function to be of
the form\cite{Chen2}
\begin{equation}
\chi(Q)=\sum_{K} G_0(Q-K)G(K) \phikq ^2
\label{chi}
\end{equation}
where $ G_0$ is the bare Green's function and $ G $ the full or dressed
Green's function which depends on the self energy $\Sigma(K)$ given by
\begin{equation}
\Sigma(K)= \sum _Q  t(Q)G_0(Q-K) \phikq ^2
\label{self-energy}
\end{equation}
Here and throughout we use the convention $\sum_Q\equiv T\sum_{i\Omega
  _n}\sum_{\bf q}$.  In the above expressions $\varphi_{\bf k}$
represents a generalized (for example) $s$- or $d$-wave function
symmetry factor.  While there are two other T-matrix approaches in the
literature, one\cite{NSR,Randeriareview} in which $\chi$ is related to
$G_0G_0$, and one which is even more widely used\cite{Haussmann} in
which $\chi$ is related to $GG$, only the $GG_0$ scheme is simply
connected to the Hartree-GL approach.\cite{Wilkins}

To compare with GL theory we expand these equations to first order in
the self energy correction\cite{schmid_comparison}. The T-matrix can be written in terms of the
attractive coupling constant $g$ as
\begin{equation}
t(Q)=  \frac{g}{1+g\chi_0(Q)+g\delta \chi(Q)}
\label{t-matrix2}
\end{equation}
where 
\begin{equation}
\chi _0(Q)=\sum_{K} G_0(Q-K)G_0(K) \phikq ^2
\label{chi2}
\end{equation}
Defining 
\begin{equation}
\Delta^2 = - \mathop{\sum_Q} t(Q)
\label{gap_approx}
\end{equation}
we arrive at (see Appendix \ref{App:weak}) 
\begin{equation}
\Sigma (K) \approx -G_0 (-K) \Delta^2 \phik ^2
\label{self_energy2_gg0}
\end{equation}
The results of Appendix \ref{App:weak} can be used to derive a self
consistency condition on $\Delta^2$ in terms of the quantity $\delta
\chi (0)$ (first order in $\Sigma$), which satisfies
\begin{equation}
\delta \chi (0)=-b_0 (\beta \Delta)^2,
\label{delta_chi2}
\end{equation}
implying that 
\begin{equation}
  \delta \chi(0)=-\frac{b_0}{N(0) T}\int \frac
  {d^3q}{(2\pi)^3}\; \frac{1}{\epsilon+\xi_1^2 q^2-\delta \chi(0)/N(0)}\;,
\label{self-cons_gg0}
\end{equation}
which coincides with the condition derived earlier in  Eq.~(\ref{gap_GL}).
Finally, defining 
\begin{equation}
\tilde {\mu}_{pair}(T)\equiv \frac{1}{N(0) t(0)}=-\epsilon+\frac{\delta\chi(0)}{N(0)}
\label{tilde_mu_pair}
\end{equation}
we may interpret the vanishing of $\tilde{\mu}_{pair}$ as the condition
that at $ T = T_c$
\begin{equation}
\tilde{\mu}_{pair}(T_c)=0.
\end{equation}
It follows from Eqs.~(\ref{delta_chi2}), (\ref{tilde_mu_pair}) and (\ref{mu_def_GL}) that 
\begin{equation}
\bar{\mu}_{pair} = \tilde{\mu}_{pair},
\end{equation}
so the above condition for $T_c$ is in agreement with that found earlier
[Eq.~(\ref{tc_cond_GL})] and the effect of a finite $\Delta(T_c)$
(self-energy correction) is a shift downward in the critical temperature
relative to its value (given by $T^*$) in the $\Delta(T_c)=0$ limit.

\section{Below $T_c$: weak coupling}
\subsection{Hartree approximated GL theory for small $\Delta(T_c)$, $T
  \approx T_c$} 

The left hand side of Eq.~(\ref{self_cons_psi_q}) may be interpreted as
the number density of bosons of momentum $q$.  Since $\bar{\mu}_{pair}
(T_c)=0$, the $q=0$ level becomes macroscopically occupied once the
system enters the superconducting region, at $T = T_c$.  To support this
assertion we investigate the behavior of $\bar{\mu}_{pair}(T)$ for
$T\approx T_c^-$. We separate out the $q=0$ term in Eq.~(\ref{gap_GL})
and write
\begin{equation}
\Delta ^2=\frac{T}{N(0) V}\:\frac{1}{-\bar{\mu}_{pair}} +\sum _{q\ne 0}
\frac{T}{N(0) V}\: \frac{1}{-\bar{\mu}_{pair} +\xi _1 ^2 q^2}\;. 
\label{eq:BEC}
\end{equation}
Eq.~(\ref{mu_def_GL}) leads to another constraint on $\Delta$, which yields
\begin{equation}
\Delta ^2= \frac{T^2}{c\,\,} \bigg(-\frac{T-T^*}{T^*} -\bar {\mu}_{pair}\bigg)\:,
\end{equation}
where $c=b_0/[N(0) V]$. Differentiating both of the above equations with
respect to $T$ one obtains the following expression for
$\frac{d\bar{\mu}_{pair}}{d T\,\,}$
\begin{eqnarray}
\lefteqn{\frac{d \bar{\mu}_{pair}}{d T\,\,}}\nonumber\\
&=&\frac{\displaystyle{\frac{\Delta ^2}{T}-\frac{T^2}{c
    T^*}}}{\displaystyle{\frac{T}{N(0) V \bar{\mu}_{pair} ^2}+\sum _{q\neq
    0}\frac{T}{N(0) V}\frac{1}{(-\bar{\mu}_{pair} +\xi _1 ^2
    q^2)^2}+\frac{T^2}{c}}} \;.\nonumber\\
\label{eq:dmu/dt}
\end{eqnarray}
The numerator of Eq.~(\ref{eq:dmu/dt}) is negative in the vicinity of
$T_c$, since $\frac{c \Delta^2}{T^2}\ll \frac{T}{T^*}$. At $T_c$, the
first term in the denominator diverges ($\frac{1}{V \bar{\mu}_{pair}}$
is a finite number in the thermodynamic limit), and as also found
elsewhere\cite{Wilkins} $\frac{d \bar{\mu}_{pair}}{dT\,\,}=0\,\,\, (T=T_c)$.
Since $\bar {\mu}_{pair}$ cannot be positive (that would make the right
hand side of Eq.~(\ref{self_cons_psi_q}) negative), and its derivative is
negative or zero, we conclude that $\bar{\mu}_{pair}$ must remain zero
in the vicinity of, but below $T_c$, where the GL description is
applicable:
\begin{equation}
\bar{\mu}_{pair}(T)=0,\quad T \approx T_c^-
\label{mu_cond_GL}
\end{equation}
 This implies, following Eq.~(\ref{mu_def_GL}),
\begin{equation}
\Delta^2(T) = - \frac{\epsilon}{b_0}\: \frac{N_0V}{\beta^2} \;.
\end{equation}
This result was previously obtained in Ref.~\onlinecite{Grossmann}.  It
is important because \textit{it shows that when the system reaches $T_c$
  the excitation gap $\Delta(T)$ assumes the BCS or mean field value}.

We notice strong analogies with Bose-Einstein condensation (BEC) in an
ideal Bose gas.  Here $\Delta^2 $ plays the role of $N$, the total
number of bosons, which below $T_c$ contains two components, one
associated with the condensate $\Delta_{q=0}^2$ and the other with the
non-condensed pairs $\Delta_{q \neq 0}^2$.  The latter are like the
excited states of the BEC system. We write
\begin{equation}
\Delta^2 = \Delta_{q=0}^2 ~ + \Delta_{q \neq 0}^2
\label{delta_breakdown_GL}
\end{equation}
It follows from Eqs.~(\ref{gap_GL}), (\ref{delta_breakdown_GL}) and
(\ref{mu_cond_GL}) that
\begin{equation}
\beta ^2 \Delta_{q \neq 0} ^2=\int \frac{d^3q}{(2 \pi)^3} \frac{1}{N_0 V
  T}\;\frac{1}{\xi_1 ^2 q^2}\;.
\label{delta_pg_tdgl}
\end{equation}
Just as in the ideal Bose gas problem, $\Delta^2(T)$ is constrained (via
the pair chemical potential condition), $\Delta_{q \neq 0}(T)$ is
constrained through the self-consistent Hartree condition and thus one
may deduce the condensate weight, or superconducting order parameter
$\Delta_{q=0}(T)$.

\subsection{Behavior near $ T = 0$ where $\Delta(T)$ is no longer small}

A useful observation can be made at this time. We have just shown that
in the Hartree theory, $\Delta (T) $ assumes the BCS value in the
vicinity of, but below $T_c$.  One expects on very general grounds that
sufficiently far away from $T_c$, pair ``fluctuation" effects are
irrelevant and that the system is described by strict BCS theory.  (In
this regime $\Delta(T)$ is no longer a small parameter).  Thus we may
infer that everywhere below $T_c$
\begin{equation}
\Delta(T) = \Delta_{BCS}(T), \quad T \leq T_c,
\label{eq:1}
\end{equation}
so that the excitation gap is given by the BCS value.\cite{Marcelja} What is
different from strict BCS theory, however, is that 
\begin{equation}
\Delta(T_c) \ne 0
\end{equation}
This is the sole effect of pair fluctuations below $T_c$. Nevertheless
it has important consequences, because it reflects the presence of 
non-condensed bosons below $T_c$, which, in turn, mirror their normal state
counterparts.
 
\subsection{T-matrix scheme below $T_c$ }

We now show that Eq.~(\ref{eq:1}) is connected to the ideal Bose gas
condition at all temperatures below $T_c$. This follows from the
analysis in Appendix \ref{App:general} in which it is shown that the BCS
gap equation is associated with a divergence of the T-matrix defined in
Eq.~(\ref{t-matrix2}), at $Q=0$ for \textit{all} temperatures below
$T_c$.  Thus $t_{pg}^{-1}(0) = 0$ implies
\begin{equation}
\mu_{pair}(T)=0,\quad T \leq T_c.
\label{eta_gg0}
\end{equation}
In order to satisfy this gap equation, together with an independent
constraint on the number of finite momentum pairs one must incorporate a
broken symmetry $\Delta_{sc} \neq 0$, which we now reformulate within
our T-matrix theory. Below $T_c$
\begin{eqnarray}
 \Sigma(K) &=&  \mathop{\sum_Q} t(Q)G_0(Q-K)\varphi_{k-q/2}^2
\label{sigma_ggo} 
\end{eqnarray}
can be decomposed into two contributions: the ``pseudogap" and
superconducting components via
\begin{eqnarray}
\label{eq:t_matrix_decomposition}
t_{pg}(Q)&=& \frac{g}{1+g\chi(Q)}, \quad Q \neq 0 \label{t-matrix_pg}\\
t_{sc}(Q)&=& -\frac{\Delta_{sc}^2}{T} \delta(Q) \label{t-matrix_sc}\\
t &=& t_{pg} + t_{sc} \label{t-matrix}
\end{eqnarray}
By defining 
\begin{equation}
\Delta_{pg}^2 \equiv - \mathop{\sum_Q} t_{pg}(Q)
\label{delta_pg}
\end{equation}
and using the divergence of $t(Q)$ at $Q=0$ to evaluate
Eq.~(\ref{sigma_ggo}), we retrieve the BCS-like  self-energy
\begin{equation}
\Sigma (K)\approx -G_0 (-K) \Delta^2 \phik ^2
\label{self_energy_gg0}
\end{equation}
with 
\begin{equation}
\Delta^2=\Delta_{sc}^2+\Delta_{pg}^2
\label{delta_breakdown}
\end{equation}
For small $\Delta(T_c)$ and $T \approx T_c$,
Eqs.~(\ref{eta_gg0}),(\ref{delta_pg}) and (\ref{delta_breakdown}) are
manifestly equivalent to their GL counterparts (with $\mu_{pair}$
corresponding to $\bar{\mu}_{pair}$\cite{bar_mu_pair}, $\Delta_{sc}$ to
$\Delta_{q=0}$ and $\Delta_{pg}$ to $\Delta_{q\neq 0}$); they give a
complete\cite{Chen2} description of the system by determining the full
excitation gap $\Delta$, order parameter $\Delta_{sc}$ and the amplitude
of propagating pairs (the pseudogap) $\Delta_{pg}$.

\section{Extended Hartree Approximation: Arbitrarily Strong Coupling $g$}

\begin{figure}
\centerline{\includegraphics[width=3.2in]{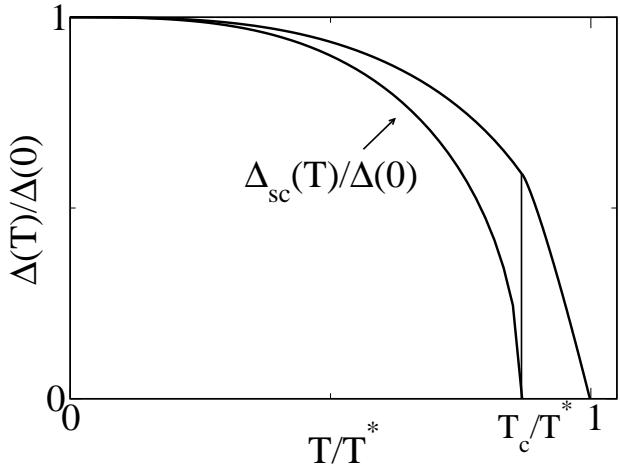}}
\caption{Schematic plot of the full excitation gap $\Delta$ and order
  parameter $\Delta_{sc}$ vs. reduced temperature $T/T^*$ for the weak
  coupling case.}
\label{fig:small_g}
\end{figure}

We have demonstrated above that there is a simple T-matrix scheme
involving the pair susceptibility ($ G G_0$) which is equivalent to
Hartree approximated GL theory both above and below $T _c$.  (Moreover,
there is a natural extension down to $ T = 0$).  This equivalence has
been proven provided we restrict ourselves to small $\Delta(T_c)$, where
GL approaches are applicable (see Fig.~\ref{fig:small_g})

\begin{figure}
  \centerline{\includegraphics[width=3.2in ]{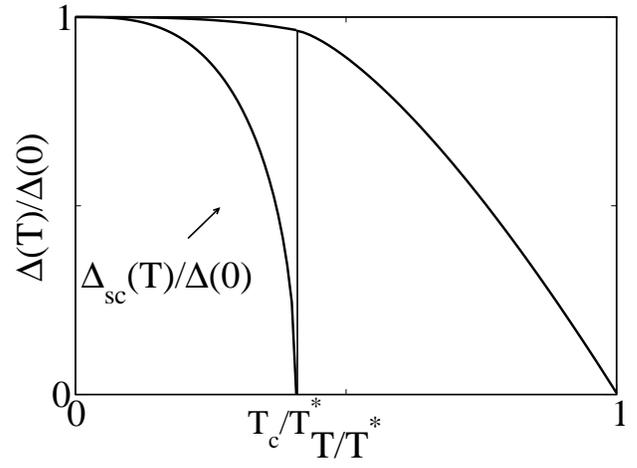}}
\caption{Schematic plot of the full excitation gap $\Delta$ and order
  parameter $\Delta_{sc}$ 
  vs. reduced temperature $T/T^*$ for the strong coupling case.} 
\label{fig:large_g}
\end{figure}

There is no reason, however, that the T-matrix scheme cannot be
considered for arbitrary coupling constant $g$, or equivalently large
$\Delta(T_c)$ where conventional GL theory is no longer appropriate.  In
this regime the separation between $T^*$ and $T_c$ can be extremely
large.  In this way, a pseudogap in the fermionic spectrum occurs at a
temperature much higher than the Bose condensation temperature $T_c$
(see Fig.~\ref{fig:large_g}). As in conventional GL theory, this
\textit{pseudogap reflects bosonic degrees of freedom}. Once bosons are
meta-stable it takes a finite excitation energy to create fermions from
them.

All the appropriate T-matrix equations have been presented above. [See
Eqs.~(\ref{chi}) and
(\ref{eq:t_matrix_decomposition})-(\ref{delta_breakdown})].  Note that
the only technical difference between the cases of weak and strong
coupling is in the details of the expression for the T-matrix itself
(see Appendix \ref{App:general}).  There is, however, an important
additional constraint which needs to be appended, on the fermionic
chemical potential $\mu$.

The resulting equations greatly simplify at and below $T_c$ because of
the vanishing of the pair chemical potential. We summarize the above
discussion by presenting expressions (appropriate to the superconducting
state) for the gap and chemical potential $\mu$
\begin{equation}                
g^{-1} +  \mathop{\sum_{\bf k}} \frac{1 - 2 f(E_{\bf k})}{2 E_{\bf
    k}}\varphi _{\bf k}^2 = 0 
\label{eq:gap_equation}
\end{equation}
\begin{equation}
n=2 \sum _{\bf k} \left[ f(E_{\bf k})+v_{\bf k}^2(1-2 f(E_{\bf k}))\right]
\label{eq:number_equation}
\end{equation}
\noindent The former follows from the vanishing of
$t_{pg}^{-1}(0)$ in Eq.~(\ref{t-matrix_pg}) and the latter from $n = 2
\sum_K G(K) $.  The quantity $v_{\bf k}$ is the coherence factor $v_{\bf
  k}^2 = \frac{1}{2}( 1 - \epsilon_{\bf k}/ E_{\bf k})$ with $\ek=k^2
/(2 m) -\mu$ and $E_{\bf k}$ is the fermionic excitation energy which
depends on the magnitude of the superconducting order parameter
$\Delta_{sc}$ and the pseudogap energy scale $\Delta_{pg}$:
\begin{equation}
E_k  = \sqrt{\ek ^2 + \Delta^2({\bf k})} 
\label{eq:Intrinsic_dispersion}
\end{equation}
where $ \Delta^2 ({\bf k}) = \Delta_{pg}^2({\bf k}) + \Delta_{sc}^2({\bf
  k}) = \Delta^2 \varphi^2 _{\bf k}$.  The important quantity
$\Delta_{pg}(T)$ here is deduced following Eq.~(\ref{delta_pg}).

\begin{figure}
\centerline{\includegraphics[width=3.2in]{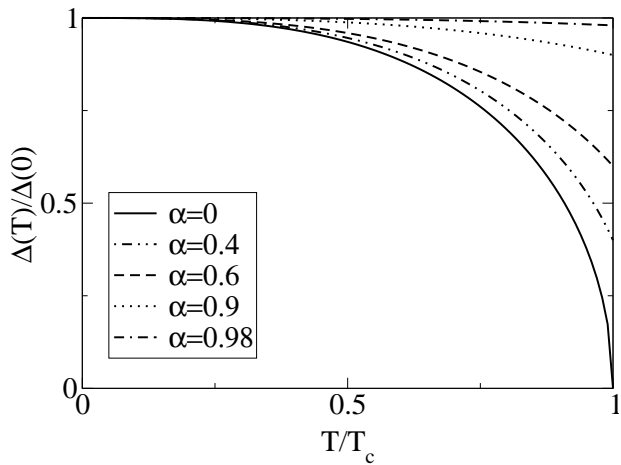}}
\caption{Implications of the ideal Bose condensation condition
  ($\mu_{pair} = 0$, when $ T \le T_c$) for the excitation gaps.  Each
  of the curves corresponds to different coupling constant strengths.}
\label{fig:schematic}
\end{figure}

At a physical level the parameter which should be associated with strong
or weak coupling is $T^*/T_c$.  However, $T^*$ is frequently difficult
to quantify. An alternative parameter for characterizing the strength of
the coupling is $ \alpha \equiv \Delta(T_c)/ \Delta(T=0)$.  This is more
readily accessed experimentally.  It should be seen from
Fig.~\ref{fig:small_g}, that the BCS limit corresponds to $
\Delta(T_c)/\Delta(0) \approx 0$, whereas (following
Fig.~\ref{fig:large_g}) in the strong coupling regime
$\Delta(T_c)/\Delta(0) \approx 1$.  Figure \ref{fig:schematic} indicates
how the temperature dependence of the gap varies with the strength of
the coupling, or alternatively with $\alpha$.  This is an important plot
because it epitomizes, perhaps, more than any other how dramatic are the
differences from strict BCS theory.  The challenge, then, is to
accommodate the results in this figure into transport and other
calculations within the superconducting phase. We thus turn to dynamical
effects associated with non-condensed bosons.
 
\section{Effective TDGL Theory for Strong Coupling Limit}

\subsection{Overview}

In this section we investigate the applicability of an effective 
time-dependent Ginzburg-Landau theory as a basis for characterizing the
\textit{dynamics} of our excitation-gap or pairing fluctuations. This
discussion is relevant to transport calculations associated with bosonic
degrees of freedom.  Quite generally, we may view TDGL as a generic
equation of motion for describing bosons interacting with a reservoir of
fermionic pairs.  This equation of motion can be associated with the
long wavelength, low frequency behavior of the pair propagator or
T-matrix, and it has a validity both above and below $T_c$, as long as
bosonic degrees of freedom are relatively stable.

In the presence of a pseudogap, the situation is complicated by the fact
that the order parameter is generally different from the excitation gap.
This is illustrated clearly in Figs.~\ref{fig:small_g} and
\ref{fig:large_g}.  Thus, below $T_c$, one has to take care to
distinguish between the fluctuations in each of the two channels. By
contrast, above $T_c$ bosonic degrees of freedom are uniquely associated
with fluctuations in the fermionic excitation gap.  To address order
parameter fluctuations in the presence of a finite gap $\Delta$ at $T_c$
has been done elsewhere\cite{Kosztin2}. It is conceptually
straightforward, but difficult to implement. Such fluctuations also
alter critical exponents within a generally narrow ``critical
fluctuation" regime.

\textit{We will not consider these order parameter fluctuations here,
  but rather presume that $\Delta_{sc}$ is spatially uniform, as in a
  mean field theoretic approach}.  The bosons of interest to us are
non-condensed pairs which in turn lead to fluctuations in the fermionic
excitation gap.  A variation in the number of pairs leads to a variation
in the excitation gap $\Delta$, simply because one has to pull apart the
constituents of the pairs to create fermions.  Moreover, our mean field
theory (associated with a BCS-like ground state\cite{Leggett}) provides
a considerable simplification, because pairing fluctuation bosons are
essentially free; they interact with fermions but not directly with each
other. Once there are boson-boson interactions (beyond mean field
theory), then necessarily there is important coupling to order parameter
fluctuations.

Fermonic degrees of freedom also contribute to transport and
thermodynamics, but there is no simple phenomenology (or counterpart of
TDGL) for addressing these terms.  When the boson contributions to
transport are small (say, as in the thermal conductivity), the fermionic
terms cannot be neglected, and these have to be computed
diagrammatically, as will be discussed in more detail below.  In other
instances the bosonic contributions are singular\cite{Varlamov_review},
or nearly so, in the vicinity of $T_c$ and TDGL calculations are
appropriate for $T_c <T <T^*$, as well as for $T \le T_c$.  In the
remainder of this section we focus only on the bosonic terms.

\subsection{TDGL above and below $T_c$ }

At a microscopic level, one can expand the T-matrix for non-condensed
pairs at small $ q , \Omega$
\begin{equation} 
t^{-1}_{pg}({\mathbf{q}}, \Omega) = a _0' \left [ \frac{a_1}{a_0'}\Omega^2 +
 (1 + i \frac{a _0''}{a _0 '}) \Omega -\frac{q^2}{2 M}+ 
\mu _{pair}\right]   \:.
\label{eq:tdgl}
\end{equation}
In order to be consistent with the linear dynamics of TDGL, the
quadratic terms in $\Omega$ are neglected in what follows.  The
coefficients in Eq.~(\ref{eq:tdgl}) are in general $T$ dependent,
although, significantly below $T^*$, the most important temperature
dependence is associated with $\mu_{pair}$, which is zero at and below
$T_c$ and negative above $T_c$.  The temperature $T^*$ corresponds to
the onset of a resonant structure in the T-matrix. This onset
temperature is reliably computed without including pseudogap effects--
thus, at the Gaussian approximation level.  Resonance
effects\cite{Janko} (reflecting the initial formation of meta-stable
pairs) enter via the ratio $a_0'/a_0''$, which (at fixed $T$ near $T^*$)
can be increased by increasing the coupling constant $g$.  The larger is
this parameter the more pronounced are pair resonances.  Stated
alternatively, as the coupling constant $g$ increases the propagating
component $a_0'$ becomes increasingly more important than the diffusive
term $a_0''$.  Indeed, similar observations were made by Randeria and
co-workers\cite{Randeriareview}.

We now proceed to the more detailed analysis of the coefficient $a_0''$,
slightly above and at all $T$ below $T_c$.  Well above $T_c$ (but below
$T^*$) more detailed numerical calculations are
required\cite{Maly1,Maly2}, and these demonstrate that $a_0''$ increases
with $T$ as pseudogap effects diminish.  At our leading order
approximation as in Eq.~(\ref{self_energy_gg0}), we do not distinguish
between lifetimes associated with the condensed and non-condensed
bosons.  For the purposes of computing $T_c$ and the various energy gaps
below $T_c$, this has been shown to be a reasonable
approximation\cite{Maly1,Maly2}, but it clearly misses important physics
associated with the onset of true phase coherence\cite{Chen4}. In this
approximation the finite momentum pairs are extremely long-lived (see
Eqs.~(\ref{chi_expr}) and (\ref{Omega_q:Gamma_q})).  Consequently,
$a_0''$ is zero. Formally, this results comes from the fact that the
imaginary part of the pair susceptibility $\Gamma (0,\Omega)$ has a
higher power than linear dependence on $\Omega$.  If, instead we
introduce (as in Appendix \ref{dirt_appendix}) a finite lifetime to
distinguish the contributions from condensed and non-condensed bosons
\begin{equation}
\Sigma(K)=\frac{\Delta _{sc} ^2 \phik ^2}{\omega +\ek}+\frac{\Delta _{pg} ^2 \phik ^2}{\omega +\ek+i \gamma}
\label{eq:sigma_gamma2}
\end{equation}
we find a non-vanishing\cite{Maly1} TDGL coefficient $a_0''$ above $T_c$.
\begin{equation}
  a_0''=\frac{N(0) \gamma}{\Delta _{pg}^2}.
\label{a_0''_gamma_text}
\end{equation}

\begin{figure}
\centerline{\includegraphics[clip,width=3.2in]{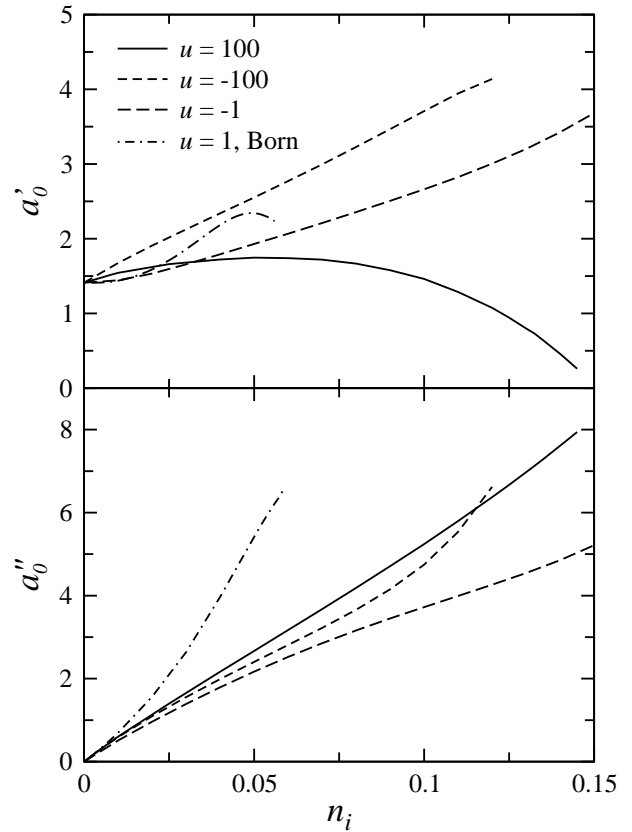}}
\caption{TDGL coefficients $a_0'$ and $a_0''$ vs. impurity concentration
  $n_i$, for several values of scattering strength $u$ from the unitary
  to the Born limit, in a self-consistent T-matrix treatment
  (Ref.~\protect\onlinecite{Chen-Schrieffer}), where the nonmagnetic,
  $s$-wave scattering potential is assumed, as given by
  $u({\mathbf{x}})=u\delta ({\mathbf{x}})$. }
\label{fig:impurity_a_0}
\end{figure}

Below $ T_c$, however, the existence of a condensate leads to $a_0''=0$,
even for the more general self energy of Eq.~(\ref{eq:sigma_gamma2}).
These arguments are presented in Appendix \ref{dirt_appendix}.  However,
the presence of even a small amount of disorder is sufficient to restore
a linear in frequency imaginary term in the TDGL expansion of the
T-matrix. These observations were made by Chen and
Schrieffer\cite{Chen-Schrieffer}.  In Fig.~\ref{fig:impurity_a_0}, we
present a figure from their work which illustrates how the frequency
dependent contributions to the T-matrix evolve with impurity
concentration, for different scattering strengths.

In this way one establishes an effective TDGL description for the
non-condensed boson dynamics both above and below $T_c$.  It should be
stressed, however, that the character of this theory changes on either
side of $T_c$.

The observation that the pairs live very much longer than anticipated
by, say, Gaussian level calculations is a consequence of the fermionic
pseudogap\cite{Maly1,Maly2}. In this way, the diffusive component (i.e.,
the parameter $a_0''$ ) remains small for an extended range of
temperatures above $T_c$, and becomes even smaller (impurity limited)
below $T_c$. This reflects the fact that as the fermions acquire a
larger gap, the bosons live longer and the two degrees of freedom become
progressively more distinct.  These same effects are underlined by our
earlier observation that $T_c$, as distinguished from $T^*$, must be
computed by including the feedback effects of the fermionic pseudogap on
the bosonic propagator. This is precisely reflected in
Eqs.~(\ref{eq:gap_equation}) and (\ref{eq:number_equation}) which,
together with Eq.~(\ref{delta_breakdown}) must be solved to determine
$T_c$.  Because of the possibility of a large separation between $T^*$
and $T_c$, (as $g$ is progressively increased) one may extend the simple
dynamics associated with TDGL theory to describe bosonic transport at
temperatures below $T^*$, \textit{not just those limited to the
  immediate vicinity of Bose condensation, $T_c$}.  This provides some
microscopic support for a recent phenomenological approach\cite{Shina}
which addresses Nernst and other normal state transport coefficients in
underdoped cuprates.

\section{Generalized Approach To Transport: T-matrix Theory Below $T_c$}

We turn now to transport properties below $T_c$, beginning with the
superfluid density. It should be clear from Figs.~\ref{fig:small_g} and
\ref{fig:large_g} that the order parameter and the excitation gap are to
be distinguished in the superconducting state.  We can thus say that the
same $T_c$ as calculated in Section \ref{sec:above_Tc_weak} via the self
consistency conditions at the (extended) Hartree-GL level will show up,
for example in $\rho_s(T)$. In particular \textit{the superfluid density
  must necessarily couple to the pair fluctuations in the
  superconducting state} in such a fashion that $\rho_s(T)$ reflects the
superconducting order parameter, rather than the fermionic excitation
gap.  This coupling of pair fluctuations to $\rho_s(T)$ can be
contrasted with the way in which collective (phase) mode contributions
enter into $\rho_s(T)$ at the BCS level. These terms are only required
to preserve general gauge invariance and these bosons do not affect
$\rho_s$ when it is computed in the transverse gauge.  The pair
fluctuations we discuss here are necessary for a consistent calculation
of the superfluid density, even in the transverse gauge.

In this section we decompose the transport contributions into two types
of excitations of the condensate: fermionic and bosonic.  It is well
known\cite{Varlamov_review} that bosonic contributions to transport
coefficients in the (less self-consistent) Gaussian-TDGL theory of
pairing fluctuations are associated with Aslamazov-Larkin diagrams. The
lowest order T-matrix theory introduces additional diagrams called the
``Maki-Thompson" and ``Density of States" contributions. These latter
two may be viewed as \textit{fermionic} contributions.

Given the self energy and form of self consistent T-matrix, Ward
identities can be used to characterize the fermionic and bosonic
contributions to transport, at the extended Hartree level.  Again,
Aslamazov-Larkin (interpreted as associated with non-condensed bosons)
and Maki-Thompson diagrams (interpreted as associated with fermionic
excitations) are present, but both contain bare as well as dressed
Green's functions.  The fact that the combination $GG_0$ appears plays
an important role, throughout our discussion. This pair susceptibility
is related to the usual Gor'kov $F$ function.

Because of the form of the pair susceptibility and the fact that below
$T_c$, $\mu_{pair} = 0$, we show below that the fermionic terms combine
to yield the usual BCS contributions to transport, which now depend on
the full gap $\Delta$, as distinct from the order parameter,
$\Delta_{sc}$.  That the fermionic contributions conspire to be of the
BCS form may seem natural at one level, but on another level, this is
highly non-trivial since at very strong coupling $\Delta$ is essentially
$T$-independent below $T_c$, as can be seen from
Fig.~\ref{fig:schematic}.  Thus, the fermionic contributions to
transport are nearly $T$ independent, in striking contrast to what is
found in the BCS or weak coupling limit.  To demonstrate how the bosonic
and fermionic contributions enter, we begin with a formulation of the
superfluid density in the presence of general self energy effects.

\subsection{Superfluid density: General formalism in transverse gauge}

\begin{figure}
  \centerline{\includegraphics[width=3.2in]{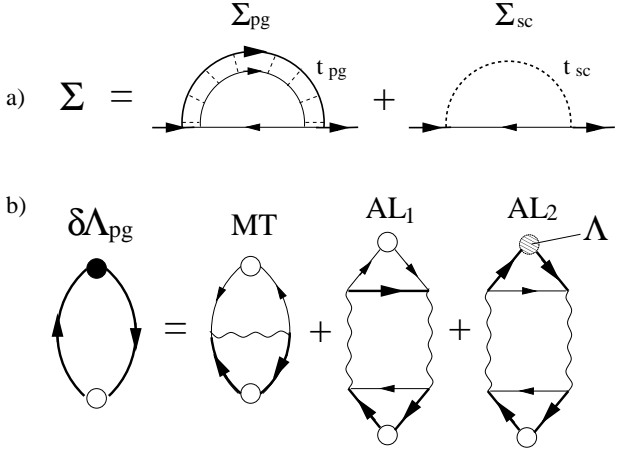} }
\caption{Self energy contributions (a) and response diagrams for 
the vertex correction corresponding to $\Sigma_{pg}$ (b).
   Heavy lines are for dressed, while light lines
  are for bare Green's functions. Wavy lines indicate $t_{pg}$.}
\label{fig:full_jj}
\end{figure}

The electromagnetic response in our extended Hartree theory is
constructed from the two additive contributions to the self energy shown
in Figure \ref{fig:full_jj}a.  The superconducting self energy
$\Sigma_{sc}$ contains an anomalous reversal of the fermion line due to
the creation of a condensed pair.  While $\Sigma_{pg}$ apparently also
has this reversed line, the charge flow is compensated by the 
fermion pair propagator, $t_{pg}$, and is thus a conventional diagram.  The
quantity $t_{pg}$, in turn, depends on one full and one bare Green's
function, each of which, as shown below, leads to additional
(Aslamazov-Larkin like) diagrams in the electromagnetic response.

We define the superfluid density, $n_s$, in terms of the magnetic London
penetration depth $\lambda_L$ as
\begin{equation}
  \lambda^{-2}_L = \frac{\mu_0 e^2 n_s}{m} \:,
\label{Lambda_Def_Eq}
\end{equation}
where $\mu_0$ is the magnetic permittivity. For convenience, we will set
$\mu_0 = e=c=1$.
Note on a lattice, $n/m$ should be replaced by 
\begin{equation}
\left(\frac{n}{m}\right)_{\mu\nu} \equiv 2\sumk \frac{\partial^2 \ek}
{\partial k_\mu \partial k_\nu} n_{\mb{k}} \:, \qquad (\mu = x,y,z) \:,
\label{Mass_Def_Eq}
\end{equation}
where $n_\mb{k}$ is the fermion density distribution in momentum space. For
the quasi-2D square lattice, the in-plane mass tensor is diagonal, and 
$(n/m)_{xx}= (n/m)_{yy}$. 

We consider the in-plane penetration depth which is expressed in terms
of the local (static) electromagnetic response kernel $K(0)$ in linear
response theory \cite{Abrikosov},
\begin{equation}
\lambda^{-2}_L=K_{xx}(0) =  \left( \frac{n}{m}\right)_{xx} - P_{xx}(0) \:,
\label{Lambda_K0_Eq}
\end{equation}
where $K$ is defined by
\begin{equation} 
J_\mu (Q) = P_{\mu\nu}A_\nu(Q) -  \left(\frac{n}
  {m}\right)_{\mu\nu}\! A_\nu(Q) = -K_{\mu\nu}(Q) A_\nu(Q) \:,
\label{K_Def}
\end{equation} 
and the current-current correlation function 
\begin{eqnarray}
\label{P_Def_Eq}
\lefteqn{P_{\mu\nu} (Q)= 
\int_0^{\beta} \:d\tau\: e^{i\Omega_n\tau} 
\langle j_{\mu}(\mb{q},\tau) j_{\nu}(-\mb{q},0)\rangle} \\
&=& -  2 \mathop{\sum_K} \Lambdak_{\mu}^{EM}(K,K+Q) G(K+Q)
\lambdak_{\nu} (K+Q,K)G(K)\:. \nonumber 
\end{eqnarray}

Here the bare vertex is given by
\begin{equation}
\bm{\lambda}(K,K+Q) = \vec{\nabla}_\mb{k}\epsilon_{\mb{k}+\mb{q}/2} =
\frac{1}{m}\left(\mb{k}+\frac{\mb{q}}{2}\right) \:,
\label{lambda_Def}
\end{equation}
where, for simplicity, we have used the dispersion for jellium in the
last step, although the generalization to the lattice is
straightforward.  The electromagnetic vertex can be written in terms of 
the corrections coming from the two self-energy components as
\begin{equation}
\bm{\Lambda}^{EM} = \bm{\lambda} + \delta\bm{\Lambda}_{pg} +
\delta\bm{\Lambda}_{sc} \:,
\label{Lambda_Eq}
\end{equation}
where $\delta\bm{\Lambda}_{pg}$ is the pseudogap term. 
For illustrative purposes we
specialize to the case $\varphi=1$ (as occurs in
$s$-wave pairing).  
The superconducting vertex contribution is given by
\begin{eqnarray}
\label{delta_Lambda_SC_Eq}
  \lefteqn{\delta\bm{\Lambda}_{sc}(K+Q,K)}\nonumber\\ &=&
  \Delta_{sc}^2
 G_0(-K-Q) G_0(-K)  \bm{\lambda}(K,K+Q)\;.\nonumber\\
\end{eqnarray}
%

At the $Q\rightarrow 0 $ limit, it satisfies
\begin{equation} 
- \delta\bm{\Lambda}_{sc}(K,K) = \frac{\partial\Sigma_{sc}(K)}{\partial
  \mb{k}} \:,
\label{Lambda_SC_Eq}
\end{equation}

This equation should be contrasted with the $ T > T_c$
Ward identity
\begin{equation} 
\delta\bm{\Lambda}_{pg}(K,K) = \frac{\partial\Sigma_{pg}(K)}{\partial
  \mb{k}} \:,
\label{Lambda_PG_Eq}
\end{equation}
The difference in sign between Eqs. (\ref{Lambda_SC_Eq}) and
(\ref{Lambda_PG_Eq}) is fundamental and arises precisely from the
anomalous nature of the $\Sigma_{sc}$ diagram.  Indeed, Eq.
(\ref{Lambda_PG_Eq}) is equivalent to the statement that above $T_c$ 
the paramagnetic and diamagnetic current
contributions to the $Q=0$ response precisely cancel. 
This cancellation appears in the superfluid density calculation
presented in the next section.
By contrast Eq.~(\ref{Lambda_SC_Eq}) expresses the
failure of this cancellation for the superconducting component, which is
naturally associated with a Meissner effect.

The vertex $\delta \Lambdak_{pg}$ may be decomposed into Maki-Thompson
($MT$) and two types of Aslamazov-Larkin ($AL_1$, $AL_2$) diagrams,
whose contribution to the response is shown here in
Fig.~\ref{fig:full_jj}b.  We write
\begin{equation}
  \delta \Lambdak_{pg} \equiv \delta \Lambdak_{MT} + \delta
  \Lambdak_{AL}^1 + \delta \Lambdak_{AL}^2 (\bm{\Lambda})\:.
\end{equation}
Note that the $AL_2$ diagram is specific to the extended Hartree scheme,
in which the field couples to the full $G$ appearing in the T-matrix
through a vertex $\bm{\Lambda}$.  It is important to distinguish the
vertex $\bm{\Lambda}$ from the full EM vertex of Eq~(\ref{Lambda_Eq}).
In particular, we write

\begin{equation}
\bm{\Lambda} = \bm{\lambda} + \delta\bm{\Lambda}_{pg} -
\delta\bm{\Lambda}_{sc} \:,
\label{Lambda_Eq2}
\end{equation}
where the sign change of the superconducting term (relative to
$\bm{\Lambda}^{EM}$) is a direct reflection of the sign 
in Eq. (\ref{Lambda_SC_Eq}).

We now note that for $\varphi=1$ there is a precise cancellation
between the $MT$ and $AL_1$ pseudogap diagrams at $Q=0$.  This follows
directly from the Ward identity
\begin{equation}
Q\cdot\lambda(K,K+Q) = G_0^{-1}(K) - G_0^{-1}(K+Q)\:, 
\end{equation}
which implies 
\begin{equation}
Q\cdot[\delta \Lambda_{AL}^1 (K,K+Q)  + \delta \Lambda_{MT}(K,K+Q)] = 0
\label{AL1_cancellation}
\end{equation}
so that $\delta \Lambdak_{AL}^1(K,K) = - \delta \Lambdak_{MT}(K,K)$ is
obtained exactly from the $Q \rightarrow 0$ limit.  

Similarly, it can be shown that 
\begin{equation}
Q\cdot\Lambda(K,K+Q) = G^{-1}(K) - G^{-1}(K+Q)
\label{GWI}
\end{equation}
The above result can be used to infer a relation analogous to
Eq.~(\ref{AL1_cancellation}) for the $AL_2$ diagram, leading to
\begin{equation}
\delta \Lambdak_{pg}(K,K) = - \delta \Lambdak_{MT} (K,K)\:,
\end{equation}
which expresses this pseudogap contribution to the vertex entirely in
terms of the Maki-Thompson diagram shown in Fig.~\ref{fig:full_jj}b.
It is evident that $\delta\bm{\Lambda}_{MT}$ is simply the pseudogap
counterpart of $\delta\bm{\Lambda}_{sc}$, satisfying
\begin{equation}
- \delta\bm{\Lambda}_{MT}(K,K) = \frac{\partial\Sigma_{pg}(K)}{\partial
  \mb{k}} \:.
\label{Lambda_MT_Eq}
\end{equation}
Therefore, one observes that for $ T \le T_c$ 
\begin{equation} 
\delta\bm{\Lambda}_{pg}(K,K) = \frac{\partial\Sigma_{pg}(K)}{\partial
  \mb{k}} \:,
\label{Lambda_PG_Eq2}
\end{equation}
which establishes that
Eq.~(\ref{Lambda_PG_Eq}) applies to the superconducting phase as well.
As expected, there is no direct Meissner contribution associated with
the pseudogap self-energy.

\subsection{Superfluid density: T-matrix approximation}

The final expression for the superfluid density is obtained by
rewriting
Eq.~(\ref{Mass_Def_Eq}) by integration by parts,
\begin{eqnarray}
\label{n_m_Eq}
\left(\frac{n}{m}\right)_{\alpha\beta} & = & 
2 \sum_{K} \frac{\partial^2
  \ek}{\partial k_\alpha \partial k_\beta } G(K) 
 =  - 2 \sum_{K} \frac{\partial \ek}{\partial k_\alpha}
\frac{\partial G(K)} {\partial k_\beta} \nonumber\\
&=& - 2 \sum_{K} G^2(K) \frac{\partial \ek} 
{\partial k_\alpha} \left( 
\frac{\partial \ek}{\partial k_\beta} + \frac{\partial \Sigma_{pg}}
{\partial k_\beta} + \frac{\partial \Sigma_{sc}}{\partial k_\beta} \right) 
\:.\nonumber\\
\end{eqnarray}
Note here the surface term vanishes in all cases.

By inserting Eqs.~(\ref{n_m_Eq}) and (\ref{P_Def_Eq}) in
Eq.~(\ref{Lambda_K0_Eq}) one can see that the pseudogap contribution to
$n_s/m$ drops out by virtue of
Eq.~(\ref{Lambda_PG_Eq2}).  The in-plane superfluid is isotropic and is
given by
\begin{equation}
\frac{n_s}{m} = 2 \sum_{K} G^2(K) \frac{\partial \ek} 
{\partial k_x} 
 \bigg[ \delta\Lambda_{sc}(K,K)_x 
- \frac{\partial \Sigma_{sc}(K)}{\partial k_x } \bigg] \:.\ 
\label{eq:ns}
\end{equation}

Equation (\ref{eq:ns}) can be readily evaluated using the
superconducting vertex and the superconducting
self-energy $\Sigma_{sc}(K) = -\Delta_{sc}^2 G_0(-K) \phik ^2$
associated with our $GG_0$-based T-matrix approach.  In addition, we
introduce an approximation in our evaluation of $G$ via
Eq.~(\ref{self_energy_gg0}), to find
\begin{eqnarray} 
\label{Lambda_General_Eq}
\frac{n_s}{m} &=&
2\sum_{\mb{k}}\frac{\Delta_{sc}^2}{\Ek^2} \left [
  \frac{1+2f(\Ek)} {2\Ek}+f^\prime(\Ek)\right] \nonumber\\
& &
\times \left [
  \left(\frac{\partial\ek}{\partial k_x}\right)^2 \!\!\phik^2 -\frac{1}{4}
  \frac{\partial \ek^2}{\partial k_x} \frac{\partial \phik^2} {\partial
    k_x}\right] \:.
\end{eqnarray}
This result includes   
the general $\varphi_{\bf k}$ factor for the 
${\bf k}$-dependence of the gap and order parameter, 
whereas it has been neglected above in Eqs. 
(\ref{delta_Lambda_SC_Eq})-(\ref{Lambda_PG_Eq2}), 
where it substantially complicates the analysis. 
 
Note that in the absence of a pseudogap (i.e., at weak coupling),
$\Delta_{sc}=\Delta$. Then Eq.~(\ref{Lambda_General_Eq}) is just the
usual BCS formula.  More generally, we can define a relationship
\begin{equation}
\left( \frac{n_s}{m} \right)  = \frac{\Delta_{sc}^2}{\Delta^2} 
\left ( \frac{n_s}{m} \right)^{BCS}  \:,
\label{Lambda_BCS_Eq}
\end{equation}
where $(n_s/m)^{BCS}$ is just $(n_s/m) $ with the overall prefactor
$\Delta_{sc}^2$ replaced with $\Delta^2$ in
Eq.~(\ref{Lambda_General_Eq}). Obviously, in the pseudogap phase,
$(n_s/m) ^{BCS}$ does not vanish at $T_c$.  That the results are so
similar to their BCS counterparts is due to our Hartree treatment of
pairing fluctuations.

Finally, from Eqs.~(\ref{delta_breakdown}) 
and (\ref{Lambda_BCS_Eq}) we may write
\begin{equation}
\left(\frac{n_s}{m}\right) = \left( 1 - \frac{\Delta_{pg}^2}{\Delta^2}
\right)
\left( \frac{n_s}{m} \right) ^{BCS} \:.
\end{equation}
Here the first term represents the contribution to $n_s$ from the usual
fermions, albeit with an unusual $T$-dependence of the gap (see
Fig.~\ref{fig:schematic}).  The second term indicates that $n_s$ is
additionally depressed by bosonic pair excitations which ensure that
$n_s$ vanishes prematurely at $T_c$, rather than $T^*$.

Finally, from Eq.~(\ref{Lambda_PG_Eq}) we can infer that the self energy
approximation of Eq.~(\ref{self_energy_gg0}) implies an approximation on
$ \delta \Lambdak_{pg}$, so that
\begin{eqnarray}
\delta \Lambdak_{MT}(K,K)  \approx  
  \Delta_{pg}^2 
G_0(-K) G_0(-K)\lambdak(K,K) \:.
  \label{MT_pg_approx2} 
\end{eqnarray}
We build on this internal consistency argument in what follows away from
$Q=0$.

\subsection{$Q \ne 0 $ electrodynamics}

The real, $\Omega \neq 0$  part of the in-plane optical conductivity can be
expressed as
\begin{equation}
\sigma(\Omega) =
\Omega^{-1} \mbox{Im}\: P_{xx}({i\Omega_n \rightarrow \Omega + i0^+})\:,
\end{equation}
which is related to the superfluid density through the f-sum rule
\begin{equation}
\frac{n_s}{m} + \frac{2}{\pi} \int^{\infty}_{0} \sigma(\Omega)\:d\Omega
= \left(\frac{n}{m}\right)_{xx} \:.
\label{sum_rule}
\end{equation}

Just as for the superfluid density, the optical conductivity involves
the same set of Maki-Thompson and Aslamazov-Larkin diagrams.  In this
section we will regroup terms so as to identify explicit fermionic and
bosonic contributions
\begin{equation}
  \bm{\Lambda}^{EM} \equiv \lambdak + \delta \Lambdak_{fermions} + \delta
  \Lambdak_{bosons} \:.
\end{equation}
Similarly, it follows that the optical conductivity contains two
contributions
\begin{equation}
 \sigma (\Omega) = \sigma ^{fermions}(\Omega) + \sigma^{bosons}(\Omega)\:, 
\end{equation}
where $\sigma ^{fermions}$ comes from the $\lambdak + \delta
\Lambdak_{fermions}$ portion of the vertex.

It is not unreasonable to take Eq.~(\ref{MT_pg_approx2}) a step further
and apply it to general $Q$, so that (below $T_c$)
\begin{eqnarray}
\lefteqn{\delta \Lambdak_{MT}(K,K+Q)}\nonumber\\ &\approx&  \Delta_{pg}^2 
 G_0(-K-Q)G_0(-K) \lambdak(K+Q,K)\;.  
\label{MT_pg_approx}
\end{eqnarray}
In effect what this approximation is saying is that for fermionic
degrees of freedom the bosons enter primarily as an
excitation gap contribution. This approximation is justified by
the same reasoning which leads to Eq.~(\ref{self_energy_gg0}), using 
the divergence of $t_{pg}(Q)$ at $Q=0$.

From Eq.~(\ref{MT_pg_approx}) and Eq.~(\ref{Lambda_SC_Eq})
it follows that the pseudogap and superconducting condensate
terms add in a natural way to introduce the full excitation gap
$\Delta$:
\begin{eqnarray}
\lefteqn{\delta \Lambdak_{fermions}(K,K+Q)}\nonumber\\ &\equiv&
 \delta \Lambdak_{MT}(K,K+Q)  + \delta \Lambdak_{sc}(K,K+Q)  \nonumber\\
&=& \Delta^2 G_0(-K-Q) \lambdak(K+Q,K)  G_0(-K) \;.
\end{eqnarray}
This term, combined with the ``density of states" contribution from
$\bm{\lambda}$, indicates that the fermionic contributions to general
transport coefficients are to be calculated within a BCS framework, but
with the full gap $\Delta$, which is to be distinguished from the order
parameter.  Though these contributions to transport are formally similar
to BCS theory, their $T$-dependence may differ considerably due to the
weak $T$-dependence of $\Delta$ in the strong pseudogap regime.

To characterize the direct contribution from bosonic degrees of freedom
to transport, which are associated with AL diagrams, we must treat their
full dynamics.  We define
\begin{eqnarray}
\lefteqn{\delta \Lambdak_{bosons}(K,K+Q)}\nonumber\\ &\equiv&
 \delta \Lambdak_{AL}^1 (K, K+Q) 
+ \delta \Lambdak_{AL}^2 (K,K+Q)
\end{eqnarray}
and  turn now to the bosonic contribution to conductivity. 

\subsection{TDGL approach to bosonic transport}

The integrated conductivity of the bosons can be deduced from the fact
that it accounts for the difference between the true superfluid density
and the BCS prediction, leading to the constraint
\begin{eqnarray}
\frac{2}{\pi} \int_{0}^{\infty}\:d\Omega\:
\sigma^{bosons}(\Omega,T)
 = \frac{\Delta_{pg}^2}{\Delta^2} 
\left(\frac{n_s}{m}\right)^{BCS}\!\!\!\!(T)\:.
\label{boson_weight}
\end{eqnarray}
The bosons make a maximum contribution at $T_c$. At this temperature,
bosons can account for as much as $90\%$ of the spectral weight at $T_c$
in a strongly pseudogapped superconductor with $T^*/T_c = 10$, while
their contribution vanishes in the weak coupling limit $T^* \rightarrow
T_c$.  The boson weight vanishes at $T=0$ at all couplings as a
consequence of the condensation of all bosonic pairs.  This pairing
fluctuation bosonic contribution is not limited to a narrow region near
$T_c$, but extends well into the superconducting state.

We define the bosonic response $P_{boson}$ as the contribution to $P$
given by $\delta \bm{\Lambda}_{bosons}$.  These terms each involve a
pair of T-matrices, and to leading order in frequency, $P_{boson}$ may
be written as
\begin{eqnarray}
\lefteqn{P_{boson}(Q)}\nonumber\\ 
&\equiv& 
-2 \mathop{\sum_K} 
\left[\delta\bm{\Lambda}_{AL}^1(K,K+Q) 
+  \delta\bm{\Lambda}_{AL}^2(K,K+Q) \right] \nonumber\\ 
&&\times {} G(K+Q) \bm{\lambda}(K+Q,K) G(K) \nonumber\\
&\approx&  \mathop{\sum_P} \bm{\Lambda}_t(P,P)
t_{pg}(P+Q)\tilde{\bm{\Lambda}}_t (P,P) t_{pg}(P)\;.
\label{boson_response}
 \end{eqnarray}
Here $\bm{\Lambda}_t$ is the vertex for $t_{pg}$ approximated as
\begin{equation}
\bm{\Lambda}_t(P,P) \approx 2 \frac{a_0'}{M} {\bf p}
\label{Lambda_t_second_approx}
\end{equation}
where we have used the T-matrix expansion of Eq.~(\ref{eq:tdgl}).
The quantity $\tilde{\bm{\Lambda}}_t = z\bm{\Lambda}_t(P,P)$ where 
$z=1$ in the normal state \cite{Patton} and is modified in the 
superconducting state. Reasonable estimates of $z(T)$ below $T_c$
may be obtained from Eq.~(\ref{sum_rule}).

If we presume Eq.~(\ref{boson_response}) holds for a range of low
frequencies we may infer a simple expression for the (in plane) ac
conductivity
\begin{eqnarray}
\label{TDGL_sigma}
\sigma^{bosons}(\Omega) &=& \frac{1}{2} z  a_0^{\prime 2} (2e)^2
 \mathop{\sum_{\bf p}} \left(\frac{p_x}{M}\right)^2 \!\int \frac{dE}{2\pi} 
\tilde{A}({\bf p},E) \nonumber\\
&&\times{}
\tilde{A}({\bf p},E+\Omega)\: \frac{b(E) - b(E + \Omega)}{\Omega}\,,
\end{eqnarray}
where we now use $\tilde{A}({\bf p},\Omega) = -2
\:\mbox{Im}\:t_{pg}({\bf p},\Omega+i0)$ for the bosonic spectral
function.  More generally, at higher frequencies the internal fermionic
structure via the $ Q $ dependence of the boson vertex must be included.
This structure results from the individual coupling of radiation to each
constituent fermion in the pair. However, it is reasonable to assume
that the compositeness of the pairs will not be resolved by radiation of
wavelengths larger than the pair size or frequencies below the
pair-breaking energy $\Delta$.  For ${\bf q = 0}$ and $\Omega < \Delta$,
then, we argue the bosonic vertex functions are well approximated by the
velocity ${\bf p}/M$ (or a constant multiple thereof.)  Indeed, for
calculations of the ac conductivity, Varlamov \cite{Varlamov2} has
argued that an analogous approximation (for Gaussian fluctuations) is
valid near $T_c$ where the pole structure of the T-matrix causes its
frequency dependence to dominate that of the Green's functions at the
vertices. At the extended Hartree level this pole structure is present
at and below $T_c$ due to the vanishing of $\mu_{pair}$. Relative to
Gaussian theory, the $Q$-dependence of the boson vertices will be
further suppressed for $\Omega < \Delta$ through the appearance of the
gapped fermion propagator $G$ in all vertex subdiagrams.

The result derived above for the bosonic contribution to the ac
conductivity is essentially the same as would obtain from ``true"
bosons, except for the constant factor $a_0^{\prime 2}$.  To see this we
review the ac conductivity in a system of free bosons of charge $e^*$,
mass $M^*$, and chemical potential $\mu^*$, in contact with a reservoir
of fermion pairs. This calculation\cite{Shina} represents a
generalization of standard TDGL-like schemes, away from the Bose
condensation temperature.  The simplest physical picture which allows
for an exactly solvable conductivity in the presence of quantum
dissipation assumes that the bosons interact with a reservoir of
\textit{localized}\cite{Shina} fermion pairs (treated as having ideal
gas Bose-Einstein statistics). This gives rise to a boson self-energy
$\Sigma_B (\Omega)$ without introducing vertex corrections to the
electromagnetic response.  Such a model yields\cite{Shina} an ac
conductivity given by
\begin{eqnarray}
\label{true_boson_sigma}
\sigma^0 (\Omega) &=& \frac{1}{2}(e^*)^2
 \mathop{\sum_{\bf p}} \left(\frac{p_x}{M^*}\right)^2 
\int \frac{dE}{2\pi} 
\tilde{A}({\bf p},E)
\tilde{A}({\bf p},E+\Omega) \nonumber\\
& &\times{} 
\frac{b(E) - b(E + \Omega)}{\Omega}\:.
 \end{eqnarray}
 Here $b(E)$ is the Bose statistical function and $\tilde{A} = -2
 \:\mbox{Im}\:B$ is the boson spectral function. The boson propagator is
 given by $B({\bf q},\Omega)^{-1} = \Omega - q^2/2M^* + \mu^* -
 \Sigma_B(\Omega)$.  The boson vertex here is the velocity ${\bf
   p}/M^*$.  The boson self-energy arises from scattering processes into
 and out of the thermal reservoir.

 To compare with the pairing theory, we note that while the boson
 self-energy in the above model arises from scattering into the
 reservoir, the self-energy of bosons in the pairing model (whose
 imaginary part we may regard as $\mbox{Im}\:t_{pg}^{-1}$) arises from
 pair dissociation and recombination processes. 
 We note, however, that fundamental differences between fermion pairs
 and true bosons remain in the analytic structure of the respective
 T-matrix and boson propagator $B$.  For true bosons, the real and
 imaginary parts of $\Sigma_B$ obey Kramers-Kronig relations and vanish
 in the high energy limit.  The propagator $B$ then reduces to its
 ``bare'' form $(\Omega - q^2/2M^* + \mu^*)^{-1}$.  The T-matrix has the
 structure $t_{pg}^{-1} = g^{-1} + \chi$ where the pairing
 susceptibility $\chi$ satisfies the same causality constraints.  The
 vanishing of $\chi$ in the high-energy limit leaves the asymptote
 $t_{pg} \rightarrow g$ and the T-matrix loses all energy and momentum
 structure due to the dissociation of all pairs.  However, this
 difference is not expected to be relevant for conductivity calculations done
 below the pair-breaking energy scale.

 We now consider the evaluation of Eq.~( \ref{TDGL_sigma}) using the
 expanded T-matrix of Eq.~(\ref{eq:tdgl}), neglecting the $\Omega^2$
 term.  Dissipation in $\sigma$ at nonzero frequencies requires $a_0''
 \ne 0$, which below $T_c$ requires impurity scattering, as discussed in
 Appendix \ref{dirt_appendix}. Here, we focus on the behavior of the
 conductivity as a function of the ratio $\nu \equiv a_0''/a_0'$, which
 in turn enters $\Gamma({\bf q},\Omega)$ in the expanded T-matrix
 approximation (Eq.~(\ref{Omega_q:exp})) as $\nu \Omega$. We note here
 that in the TDGL approach for Gaussian fluctuations near $T_c$, this
 ratio is typically large (of order $(E_F / T_c)$) due to the fact that
 $a_0'$ is a measure of particle-hole asymmetry. As shown earlier in
 Fig.~\ref{fig:impurity_a_0}, $\nu$ tends in the extended Hartree theory
 to increase rapidly from zero, becoming of order unity with the
 introduction of a modest concentration of impurities.
 
TDGL calculations are typically done in the classical regime $\Omega \ll
T$, which allows the simplification $[b(E) - b(E+\Omega)]/\Omega
\rightarrow T/[E(E+\Omega)]$ in Eq.~(\ref{TDGL_sigma}). The conductivity
below $T_c$ is then found to have the characteristic
$\Omega^{-1/2}$ dependence\cite{Varlamov2} for all values of $\nu$:
\begin{equation}
\sigma^{boson}(\Omega) \rightarrow \frac{2}{3 \pi} z e^2 T \left(\frac{2M}{\Omega}\right)^{1/2}
  \left(\frac{1+\nu^2}{\nu}\right)^{1/2} .
\end{equation}
To compute the conductivity in the superconducting state up to $\Omega
\sim \Delta$, however, requires the inclusion of quantum statistical
factors when $\Delta \gg T_c$, in which case the $\nu$ parameter affects
the frequency-dependence of $\sigma^{boson}$.  While the low-frequency
limiting behavior is $\Omega^{-1/2}$ independent of $\nu$, $\sigma$
falls faster than $\Omega^{-1/2}$ outside the classical regime before
crossing over to $\Omega^{1/2}$ behavior at higher frequencies.  This
crossover may extend over a large frequency range, depending on the
value of $\nu$, so that generally $\sigma$ is characterized by
$\Omega^{-1/2}$ at low frequencies and a long high frequency tail which
cannot be integrated to infinity.  Since the integrated bosonic weight
is finite, it is reasonable to expect that this expression for the
conductivity is cut off above the pair-breaking scale (at several times
$\Delta(0)$), where the TDGL formulation is known to break down.

\section{Conclusions}

This paper is based on the observation that the BCS ground state
wavefunction has a more general validity\cite{Leggett,Eagles}.  By
increasing the strength of the attractive interactions,
superconductivity in this state progressively takes on the character of
Bose Einstein condensation.  That the same $T=0$ wavefunction can apply
to a system where pair formation and pair condensation are associated
with different energy scales provides support for the notion that there
exists a mean field theory of a more general nature than that of strict
BCS theory.  In the more general case the various energies of the BCS
picture are no longer degenerate: $T^* \ne T_c$ and $\Delta \ne
\Delta_{sc}$.

In this paper we have shown that this generalization of BCS physics is
to be found in a treatment\cite{Wilkins,Patton1971} of pair fluctuation
effects at the Hartree level.  This establishes that the excitation gap
$\Delta(T_c)$ is non-zero, or equivalently that there is a pseudogap,
even in the weak coupling limit, as was observed experimentally many
years ago\cite{Abeles}.  Going beyond this previous work and into the
broken symmetry phase, a self consistent analysis leads to the condition
that the pair chemical potential $\mu_{pair} = 0 $ for $T \le T_c$, and
that this is equivalent to the statement that the excitation gap
$\Delta(T)$, assumes the BCS value at and below $T_c$.  It follows from
the fact that the order parameter is necessarily zero at $T_c$, that the
excitation gap and the superconducting order parameter are
distinguishable below $T_c$.  This behavior is schematically illustrated
in Figs.~\ref{fig:small_g} and \ref{fig:large_g}.

Because it exists in conjunction with fermionic degrees of freedom, the
ideal Bose character found here supports true superconductivity. This is
demonstrated here by  calculations of the superfluid density.
More generally, transport properties contain two types of contributions
from both fermionic and bosonic excitations.   At
finite temperatures, the fermionic terms are well approximated by the
usual BCS contributions to transport \textit{but with a highly-non
  -BCS-like, often temperature independent gap, $\Delta(T)$}.

At this extended Hartree level, fluctuations in the order parameter
$\Delta_{sc}$ and in the excitation gap $\Delta$ represent
distinct channels for bosonic effects in the superconducting state.  It
is more convenient to decompose these into ``condensed" and
``non-condensed" bosons.  The latter are the focus of the present paper,
and we refer to these as ``pair fluctuations" as distinct from
fluctuations of the order parameter.  Order parameter fluctuations were
discussed in the context of our extended Hartree theory in earlier
work\cite{Kosztin2}.  Non-condensed bosons are present above and below
$T_c$, but absent (like fermionic excitations) at $T=0$.  These long
lived, metastable states, even above $T_c$ (albeit, below $T^*$) are
associated with our extended Hartree treatment which introduces a
pseudogap in the fermionic spectrum.  This ``gap'' then inhibits
dissociation of the bosons into fermions.

The dynamics of these non-condensed bosons is reasonably described by
time-dependent Ginzburg-Landau theory.  While the microscopic character
of this TDGL changes abruptly from above to below $T_c$, this approach
offers a very powerful technique\cite{Shina,Iyengar} for addressing the
pair fluctuation or bosonic contributions to transport.  In this paper
we have applied these results to calculations of the ac conductivity.
One could equally well address the thermal conductivity and we can
anticipate some of the ensuing conclusions.  When both bosonic and
fermionic excitations are present (i.e., away from $T=0$), we expect
that the Wiedemann-Franz law is violated.  Because of the associated
soft energy scale, bosonic contributions to the thermal conductivity are
considerably smaller than their counterparts for the electrical
conductivity.  In this way the thermal conductivity $\kappa$ in the
superconducting state should be well approximated by considering only
BCS- like contributions, but with the anomalous temperature dependent
excitation gap shown in Fig.~\ref{fig:schematic}. Moreover, because the
bosonic contributions are unimportant in $\kappa$, we expect to recover
the well-known universal result\cite{Lee1993} for this property. This is not
necessarily true for the electrical conductivity.

In this paper we have not made reference to specific physical systems
where our mean field theory may have some applicability.  In addition to
short coherence length superconductors, (among these the high
temperature cuprates), the present picture may also be relevant
\cite{Griffin,Holland,Jin} to fermionic superfluidity in atomic trap
experiments.

\acknowledgments

This work was supported by NSF-MRSEC Grant No.DMR-0213745 (JS, AI and
KL), NSF grant No.~DMR0094981 and JHU-TIPAC (QC). We thank Shina Tan, I.
Ussishkin and A.  Varlamov for useful conversations.

\appendix
\section{T-matrix for small $\Delta$ at weak coupling}
\label{App:weak}

If we replace the dressed Green's function in Eq.~(\ref{chi}) by the bare
Green's function $G_0$, we obtain
\begin{equation}
t_0({\bf q},\Omega_n)=\frac{g}{1+g\chi _0(Q)}=-\frac{1}{N(0)}\:\frac{1}{\epsilon+a|\Omega_n|+\xi_1^2 q^2}\:.
\end{equation}
The self energy $\Sigma(K)$ is then given by
\begin{equation}
\Sigma(K)=\sum _Q  t(Q)G_0(Q-K) \phikq ^2
\label{ap_self-energy}
\end{equation}
To calculate the pair susceptibility to the first order in self energy,
we write the dressed Green's function in Eq.~(\ref{chi}) as
$G=G_0+G_0\Sigma G_0$, which gives the first order correction
\begin{equation}
\delta\chi(Q)=\sum _K G_0(Q-K) G_0(K)\Sigma (K)G_0(K)
\label{delta_chi}
\end{equation}
Since  $t(Q)$ is sharply peaked around $Q=0$  we disregard the
momentum dependence of  $G_0$ in Eq.~(\ref{ap_self-energy}), and obtain
\begin{equation}
\Sigma(P)\approx -G_0 (-K) \Delta ^2 \phik ^2
\end{equation}
where 
\begin{equation}
\Delta ^2=-\int \frac{d^3q}{(2\pi)^3}t({\bf q},\omega_n=0)
\end{equation}
Putting this back into Eq.~(\ref{delta_chi}) we find 
\begin{equation}
\delta \chi(0)=-b_0 \beta ^2 \Delta ^2 =b_0/T^2 \int
\frac{d^3q}{2\pi)^3} t({\bf q},\omega _n=0) 
\end{equation}
where $b_0/T^2=\sum _K G_0^2(K)G_0^2(-K) \phik ^2=N(0)
\frac{7\xi(3)}{8(\pi T)^2}$ (this last equality holds only for $s$-wave
pairing).  Using the expression for the T-matrix corrected by $\delta
\chi (0)$ via Eq.~(\ref{t-matrix2})
\begin{equation}
t(Q)=-\frac{1}{N(0)}\:\frac{1}{\epsilon+a|\Omega_n|+\xi_1^2
  q^2-\delta \chi(0)/N(0)}, 
\end{equation}
the self-consistency equation reads
\begin{equation}
\delta \chi(0)=-\frac{b_0}{N(0) T}\int \frac
{d^3q}{(2\pi)^3}\:\frac{1}{\epsilon+\xi_1^2 q^2-\delta \chi(0)/N(0)} \:.
\label{ap_self-cons_gg0}
\end{equation}

\section{T-matrix for arbitrary $\Delta$, below $T_c$ for general coupling}
\label{App:general}

From Eq.~(\ref{self_energy_gg0}) and $G_0(P)=1/(i\omega_n-\epsilon_{\bf
  p})$ we find $G(P)=(i\omega _n+\epsilon _{\bf
  p})/[(i\omega_n)^2-E_{\bf p}^2]$.  Here $\epsilon _{\bf p}$ is the
electron normal state dispersion measured with respect to the chemical
potential $\mu$, while $E_{\bf p}= \sqrt{\epsilon_{\bf p}^2+\Delta^2
  \varphi_p ^2}$.  Thus after performing the Matsubara sum and
analytical continuation $i \Omega _n \rightarrow \Omega +i 0^+$,
Eq.~(\ref{chi}) becomes
\begin{eqnarray}
 \chi ({\bf q},\Omega)&=& \sum_{\bf k} 
\left[ \frac{1-f(E_{\bf
     k})-f(\epsilon_{\bf k-q})} 
{E_{\bf k}+\epsilon_{\bf k-q}-\Omega -i 0^+}u_{\bf k}^2 
\right. \nonumber \\
&&{} \left. -\: \frac{f(E_{\bf k})-
f(\epsilon_{\bf k-q})}{E_{\bf k}-\epsilon_{\bf k-q}+\Omega +i 0^+}v_{\bf
     k}^2 \right] \phikq ^2  \:.\hspace{2ex}
\label{chi_expr}
\end{eqnarray}

In the long wavelength, low frequency limit, one can expand the inverse $T$
matrix as:
\begin{eqnarray} 
t^{-1}_{pg}({\mathbf{q}}, \Omega) &=& g^{-1}+\chi ({\bf q},\Omega)
\nonumber \\ 
&= &  a_1\Omega^2 + a_0'(\Omega - \frac{q^2}{2 M}
+ \mu _{pair}  +i \Gamma_{{\mathbf{q}}, \Omega})\:.\hspace{5ex}
\label{Omega_q:exp}
\end{eqnarray}

The linear contribution (\textbf{q}) is absent due to the inversion
symmetry (${\bf q} \leftrightarrow - {\bf q}$) of the system.

In the weak coupling limit, the ratio $a_0'/a_1$ is vanishingly small; when
the system has exact particle-hole symmetry (e.g., a 2D tight binding band
at half filling with a nearest neighbor hopping), $a_0'$ vanishes. In this
case the dispersion determined via
\begin{equation}
t^{-1}_{pg}({\mathbf{q}}, \Omega) = 0
\end{equation}
is linear in $q$, $\Omega_{\bf q} \sim cq$, which shows up in the
dispersion, \textit{only in the very weak coupling limit} or where there
is exact particle-hole symmetry.

In the absence of particle-hole symmetry, as $g$ increases, $a_0'/a_1$
increases, thus $a_0' \Omega$ gradually dominates and we find the
important result: $\Omega_{\bf q} \sim q^2$. For any finite $g$ and
arbitrarily small $q$, the dispersion is always quadratic, at the lowest
energies.

We are interested in the moderate and strong coupling cases, where we can
drop the $a_1 \Omega^2$ term in Eq.~(\ref{Omega_q:exp}), and hence we have
\begin{equation} 
t_{pg}({\mathbf{q}}, \Omega) = \frac{a_0^{\prime -1}}{\Omega -
  \Omega_{\mathbf{q}} 
  + \mu_{pair} + i\Gamma_{{\mathbf{q}}, \Omega}}\;,
\label{Omega_q:t}
\end{equation}
where 
\begin{equation}
\Omega_{\mathbf{q}} \equiv \frac{q^2} {2 M} 
\end{equation}
is quadratic. This defines the effective pair mass, $M$. Below $T_c$, we
have 
\begin{equation}
\mu _{pair}(T)=0,\quad T\leq T_c
\end{equation}
Using Eq.~(\ref{chi_expr}) this condition can be shown to be equivalent to
Eq.~(\ref{eq:gap_equation}), the familiar BCS gap equation. 

\section{Imaginary part of the inverse T-matrix -- bosonic lifetime
near $T_c$ and below}
\label{dirt_appendix}

In this Appendix, for definiteness, we assume $d-$wave pairing so that
$\phik =2 \cos (2 \phi)$, where $\phi $ is the polar angle.
By taking the imaginary part of the inverse T-matrix
using Eq.~(\ref{chi_expr}) we obtain the expression
 \begin{eqnarray}
 \Gamma_{\mb{q},\Omega} &=& \frac{\pi}{a_0'}\! \sumk 
\Big[
 [1-\!f(\Ek)-\!f(\ekq)] \uk^2 \delta (\Ek+\ekq-\Omega)
  \nonumber\\ 
&&{}+[f(\Ek)-\!f(\ekq)] \vk^2  \delta(\Ek\!-\ekq+\Omega)
 \Big] \phikq^2  \:.\nonumber\\
\label{Omega_q:Gamma_q}
\end{eqnarray}
For small $\Omega \ll T$, and setting ${\bf q}=0$  we can expand the
Fermi functions to first order in $\Omega$: 
 \begin{eqnarray}
\Gamma_{0,\Omega} &=&-\frac{\pi}{a_0'} \Omega \sumk \Big[
  f'(\Ek) \uk^2 \delta (\Ek+\ek-\Omega) \nonumber \\
&&{} +f'(\Ek) \vk^2 \delta(\Ek-\ek+\Omega)  \Big]\phik ^2
 \nonumber \\ &=&\frac{a_0''}{a_0'} \Omega   \:.
\label{eq:Gamma_expansion}
\end{eqnarray}
Here $\Gamma (0,\Omega )$ reflects the rate of decay of non-condensed
bosons into a bare and dressed fermion.  The $\delta$ functions in the
above expression determine the energies of these fermions $\ek$ and
$\Ek$ to be of order $\frac{\Omega ^2+\Delta ^2 \phik ^2}{\Omega}$. At
temperatures of interest ($T<T_c$ and $T \gtrsim T_c$), away from the
nodes (where $\phik =0$) and for $\Omega < \Delta $ this energy is
large. The only appreciable contribution then comes from parts of the
Fermi surface near the nodes (of angular width of the order of
$\sqrt{\frac{\Omega T}{\Delta ^2}}$), giving a higher power than linear
in $\Omega$ dependence of $\Gamma (0,\Omega)$ and $a_0''=0$.

This result is improved upon by including the effects of fermion-boson
scattering. Quite generally in the absence of impurity scattering one
finds that
\begin{equation}
a_0''=-\frac{1}{2} \sum _{\bf k} \phik ^2 A({\bf k},-\ek) f'(\ek)
\label{eq:a_0''_spectral}
\end{equation} 
where $A({\bf k},E)=$-2 \mbox{Im} $G({\bf k},i\omega _n \rightarrow E +i 0^+)$
is the fermion spectral function. Fermion-boson scattering broadens the
spectral function and $A({\bf k},-\ek) f'(\ek)$ is generally non-zero in
the normal state for all ${\bf k}$, implying that $a_0'' \neq 0$. One
way to implement this is to introduce an inverse lifetime $\gamma$ into
the expression for the pseudogap self-energy:
\begin{equation}
\Sigma(K)=\frac{\Delta _{sc} ^2 \phik ^2}{\omega +\ek}+\frac{\Delta
  _{pg} ^2 \phik ^2}{\omega +\ek+i \gamma} 
\label{eq:sigma_gamma}
\end{equation}
leading to the following expression for the spectral function $A({\bf
  k},\omega)$, above $T_c$
\begin{equation}
A({\bf k},\omega)=\frac{2 \Delta_{pg}^2 \phik ^2 \gamma}{(\omega ^2-\Ek
  ^2)^2+\gamma ^2 (\omega -\ek )^2} 
\label{eq:spectral_above_Tc}
\end{equation}
As is evident from Eq.~(\ref{eq:spectral_above_Tc}) , the quasiparticle
peaks are broadened as a result of a non-zero $\gamma$.  Thus, above
$T_c$ the imaginary part of the inverse T-matrix (or, equivalently, of
the pair susceptibility) is now
\begin{eqnarray}
\label{eq:Im_chi_gamma}
&& \mbox{Im} \chi({\bf q},\Omega)= \Gamma ({\bf q},\Omega)= \gamma \Delta
_{pg} ^2 \\ 
&& \times{}\sum _{\bf k} \frac{1-f(\Omega -\ekq)-f(\ekq)}{[(\Omega
 \! -\ekq)^2\!-\!\Ek^2]^2+\gamma ^2(\Omega\! -\!\ekq\!-\!\ek)^2} \phik^2
\phikq^2 \:.
\nonumber 
 \end{eqnarray}
 Since the quasiparticle peaks are broadened, the boson can now decay
 into states near the Fermi surface not only near the nodes but
 everywhere else on the Fermi surface.  $\Gamma (0,\Omega)$ is now
 linear in $\Omega$ and the coefficient of proportionality is easily
 found to be
\begin{equation}
\frac{\partial}{\partial \Omega} {\rm Im} t^{-1} (0,\Omega)\Big\vert
_{\Omega =0}=-\frac{\gamma \Delta _{pg} ^2}{T} \sum _{\bf k} \frac
{f'(\ek) \phik ^4}{\Delta _{pg} ^4 \phik ^4+4 \gamma ^2 \ek ^2}. 
\label{a_0''_gamma}
\end{equation}
The second term in the denominator of the summand in the above
expression can be safely neglected if $\gamma $ is taken to be much
smaller than $\Delta _{pg}$, since the main contribution to the sum
comes from the vicinity of the Fermi surface (where $\ek$ is small). The
only place where this term is not negligible in comparison with $\Delta
_{pg}^4 \phik ^4$ is in a small region near the nodes (angular width
$\propto \frac{\sqrt{\gamma T}}{\Delta _{pg}}$). However, this
correction can also be neglected, and therefore
\begin{equation}
a _0''=\frac{\gamma}{4 T \Delta _{pg} ^2 } \sum _{\bf k} \frac{1}{
    \cosh ^2 \frac{\ek}{2 T}}\:, 
\end{equation}
which, assuming that $T \ll \mu$, after a standard procedure gives, as
expected, a non-zero result 
\begin{equation}
a_0''=\frac{N(0) \gamma}{\Delta _{pg}^2}\:.
\label{eq:a_0''_gamma_result}
\end{equation} 

Below $T_c$ however $A({\bf k},-\ek)$ in Eq.~(\ref{eq:a_0''_spectral})
vanishes due to the sharpness of the superconducting self-energy, as is
easily checked by inspecting the expression for the spectral function
below $T_c$ following from the self-energy Eq.~(\ref{eq:sigma_gamma})
\begin{equation}
A({\bf k},\omega)=
\frac{2 \Delta _{pg}^2 \phik ^2 \gamma (\omega +\ek)^2}{(\omega +\ek)^2
  (\omega ^2-\Ek ^2)^2+\gamma ^2 (\omega ^2-\ek ^2-\Delta _{sc} ^2 \phik
  ^2)} \:.
\label{eq:spectral_below_Tc}
\end{equation}
Thus we can infer that $a_0''=0$.  Nevertheless, a small amount of
impurity scattering will restore a non-vanishing $a_0''$, since it is
likely to produce a non-zero $A({\bf k},-\ek)$ in
Eq.~(\ref{eq:a_0''_spectral}).  In the $d-$wave case, the impurity
renormalized imaginary part of the pair susceptibility has the following
form \cite{Chen-Schrieffer}
\begin{eqnarray}
\lefteqn{\chi''({\bf q},\Omega +i 0^+)=-\sum _{\bf k} \int _{-\infty} ^{\infty}
\frac{d\omega}{2 \pi}\: {\rm Im}\: G^R (\omega, {\bf k}) }\nonumber\\
&&{}\times A_0(\Omega -\omega,{\bf q}-{\bf k})
  [ f(\Omega -\omega)-f(\omega)] \phikq ^2    \:,\hspace{4ex}
\end{eqnarray} 
where both ${\rm Im} G^R$ (the spectral function of the full Green's
function $G$) and $A_0$ (the spectral function of the bare Green's
function $G_0$) are dressed with the impurity self-energy.  To see the
order of magnitude of this effect, we make a crude approximation of this
formula by assuming that only the full $G$ is dressed by impurities
but not $G_0$
which is equivalent to the statement that expression
(\ref{eq:a_0''_spectral}) with renormalized $A$ is valid.  We model the
impurity self-energy in the Born limit by $\Sigma_{imp}(\omega)=-i s
\vert \omega \vert$ (as is reasonable close to the Fermi surface in the
$d-$wave case) , where $s$ is a dimensionless constant parameterizing
the concentration of scatterers. Assuming $s \ll 1$ (clean limit), we
obtain
\begin{equation}
  A ({\bf k},-\ek)=-2 \frac{s \vert \ek \vert \Delta ^2 \phik ^2}{\Delta
    ^4 \phik ^4+4 s^2 \ek ^4}
\end{equation}
which, in conjunction with Eq.~(\ref{eq:a_0''_spectral}) and identifying
$-s \vert \ek \vert$ with $\gamma$ yields an expression identical to
Eq.~(\ref{a_0''_gamma}). Using reasoning similar to that leading to
Eq.~(\ref{eq:a_0''_gamma_result}), we arrive at the following estimate
for $a_0''$ in the presence of impurities:
\begin{equation}
a_0''=2 \ln 2\: N(0) \frac{s T}{\Delta ^2}\:.
\label{eq:a_0''_dirt_result} 
\end{equation}     
More complete numerical results are shown in Fig.~\ref{fig:impurity_a_0}.


\begin{thebibliography}{10}
\bibitem{bar_mu_pair}
Note here that $\bar {\mu}_{pair}$ is a dimensionless number, while $\mu
  _{pair}$ has the dimension of energy. These two are proportional in the weak
  coupling limit.
\end{thebibliography}

\end{document}